\documentclass[journal abbreviation, manuscript]{copernicus}
\nolinenumbers 

\begin{document}

\title{Combining a ground-based UV network with satellite maps: A case study for Germany}

\Author[1][barbara.klotz@i-med.ac.at]{Barbara}{Klotz}
\Author[1]{Verena}{Schenzinger}
\Author[1]{Michael}{Schwarzmann}
\Author[1,2]{Axel}{Kreuter}

\affil[1]{Institute of Biomedical Physics, Medical University Innsbruck, Innsbruck, Austria}
\affil[2]{LuftBlick OG, Innsbruck, Austria}

\runningtitle{UV Index maps of Germany}

\runningauthor{Barbara Klotz}

\firstpage{1}

\maketitle

\begin{abstract}
A study for the comprehensive information of current UV exposure for the area of Germany, based on the method for near real time calculation of UV Index maps used in the framework of the Austrian UV Monitoring Network, is presented. For the area of Germany about 22.000 surface UV Index maps were calculated for the year 2022 via the radiative transfer model libRadtran by incorporating daily forecast data for ozone, albedo and aerosols from the Copernicus Atmospheric Monitoring Service and taking into account cloud information gathered from SEVIRI data of Meteosat Second Generation in the form of a cloud modification factor. Ground-based measurements of 17 stations of the German Solar Monitoring Network were then compared to the modelled maps. For most stations the correlation coefficient between measured and modelled UV Index (UVI) is above 0.92 and the mean difference of modelled UVI to measured UVI is smaller than 0.5~UVI. The modelling of the UVI at the high mountain station Schneefernerhaus is associated with higher uncertainties (correlation coefficient 0.85 and mean UVI difference 0.6 UVI) due to the small-scale topography with spatially highly variable albedo and clouds. 
Moreover, case studies for specific locations with respect to cloud conditions and topography are discussed, as well as a case study of the combination of ground-based measurements and modelled UVI maps in form of spatial correction factors.
\end{abstract}

\introduction  
Depending on the level of exposure, solar ultraviolet (UV) radiation has beneficial as well as damaging effects on human health. Moderate exposure has a positive impact on blood pressure and mental health (\cite{Juzeniene2011}) and it is the main driver for the synthesis of vitamin D in human skin (\cite{webb2006}, \cite{Webb2011}), which is associated with an improved immune function. Overexposure to UV radiation on the other hand can lead to adverse effects like sunburn (erythema), eye damage (photokeratitis, cataract) and skin cancer (\cite{Lucas2006}, \cite{Juzeniene2011}).  

With the discovery of the Antarctic ozone hole in 1985, growing concerns about rising ultraviolet (UV) levels on the earth's surface led to the establishment of monitoring networks.
Although ground-based measurements provide the actual radiation levels on the earth's surface, their spatial range of validity is confined to the surrounding of the measurement station since local conditions of cloud cover, ozone, albedo and aerosol amount have a strong impact on the measured radiation. On the other hand, satellite-based imagery offers comprehensive coverage of geographic areas. However, their acquisition is not continuous, and the granularity of the resulting data is depending on the spatial, spectral and radiometric resolution of the measurement device onboard the satellite. Moreover, with this technique it is not possible to  measure the UV radiation at the earth's surface directly, but there are methods of combining radiative transfer modelling with satellite data to calculate UV maps (\cite{Verdebout2000}, \cite{Schallhart08}, \cite{Kosmopoulos21}, \cite{amt-2023-188}).

Nowadays the method described in \cite{amt-2023-188} is in routine use in the framework of the Austrian UV Monitoring Network to calculate UV Index (UVI, see definition in \cite{who2002}) maps of Europe every 15 minutes (see \url{ www.uv-index.at}).
In the present study the same method is applied to the area of Germany and the calculated UV Index maps are compared to measurements of 17 stations of the German Solar Monitoring Network for the year 2022. The peculiarities of the statistics of the UVI difference of modelled to measured UVI for clear, cloudy and all sky  conditions are discussed and case studies of specific locations with respect to cloud conditions and topography are presented. For cloudy conditions as well as locations with highly variable topography (station Schneefernerhaus) the agreement between model and measurement deteriorates in comparison to the clear sky case.
These features can also be seen with the UVIOS model described in \cite{Kosmopoulos21}, to which the results of the present study are compared. 

To enhance the quality of the modelled UV Index maps, the objective is to explore opportunities for combining ground-based measurements with satellite-derived maps, thereby benefiting of the high accuracy of ground measurements alongside the extensive coverage provided by satellite data. A case study of a correction factor map, gained from the ratio of measured to modelled UV Index, is discussed, highlighting the problems involved in combining both datasets.  

\section{Method}
The procedure of producing UV Index maps used for this study is described and discussed in detail in \citet{amt-2023-188}, so here a summary is provided.

First a lookup table (LUT) of erythemal irradiance at the surface for clear sky conditions is pre-calculated with the sdisort solver of the radiative transfer model libRadtran (\cite{libradtran1}, \cite{libradtran2}). The entries of the LUT are solar elevation, height above sea level (\unit{m}~a.s.l.), total ozone column, albedo and Angstrom $\beta$.   

Then clear sky UV Index maps for the area of Germany are extracted from the LUT by providing day-to-day forecasts of ozone, albedo and aerosol optical depth (AOD at 550~\unit{nm}) from the Copernicus Atmospheric Monitoring Service (CAMS) global model forecast (\citet{CAMSdata}), where Angstrom $\beta$ is calculated from the AOD with Angstrom α~1.4. For the spatial distribution of the height a.s.l. the digital elevation model GTOPO30 (\citet{GeschEA99}) is used and the solar elevation is calculated to match the timing of satellite cloud information from the SEVIRI instrument onboard the MSG (Meteosat Second Generation) satellite.

In order to adapt the clear sky model maps to cloudy conditions, SEVIRI imagery is utilized to derive cloud modification factors (CMF) for the area of Germany every 15 minutes. 
The process of calculating a CMF is only done for pixels that are considered to be cloudy by the cloud mask obtained from the MSG Meteorological Products Extraction Facility (\cite{clm}). Then satellite images at 600~\unit{nm} recorded by the SEVIRI instrument (\cite{seviri}) together with a very simple min-max scaling of the irradiance at the top of the atmosphere are used to estimate the CMF, which is a number between zero and one (zero meaning that there is an overcast sky with no radiation passing through and one describing a cloudless sky). 
Finally, the clear sky model maps are scaled with the corresponding CMF to obtain near real time UV Index maps of Germany. 

For this study almost 22.000 UV Index maps were calculated for various solar elevations and throughout the entire year of 2022, by incorporating CAMS forecasted data and satellite cloud information. Then the modelled values  were compared with ground-based measurements of 17 sites of the German Solar Monitoring Network. 

\subsection{Ground measurements}
\begin{figure*}[t]   
	\includegraphics[width=12.0cm]{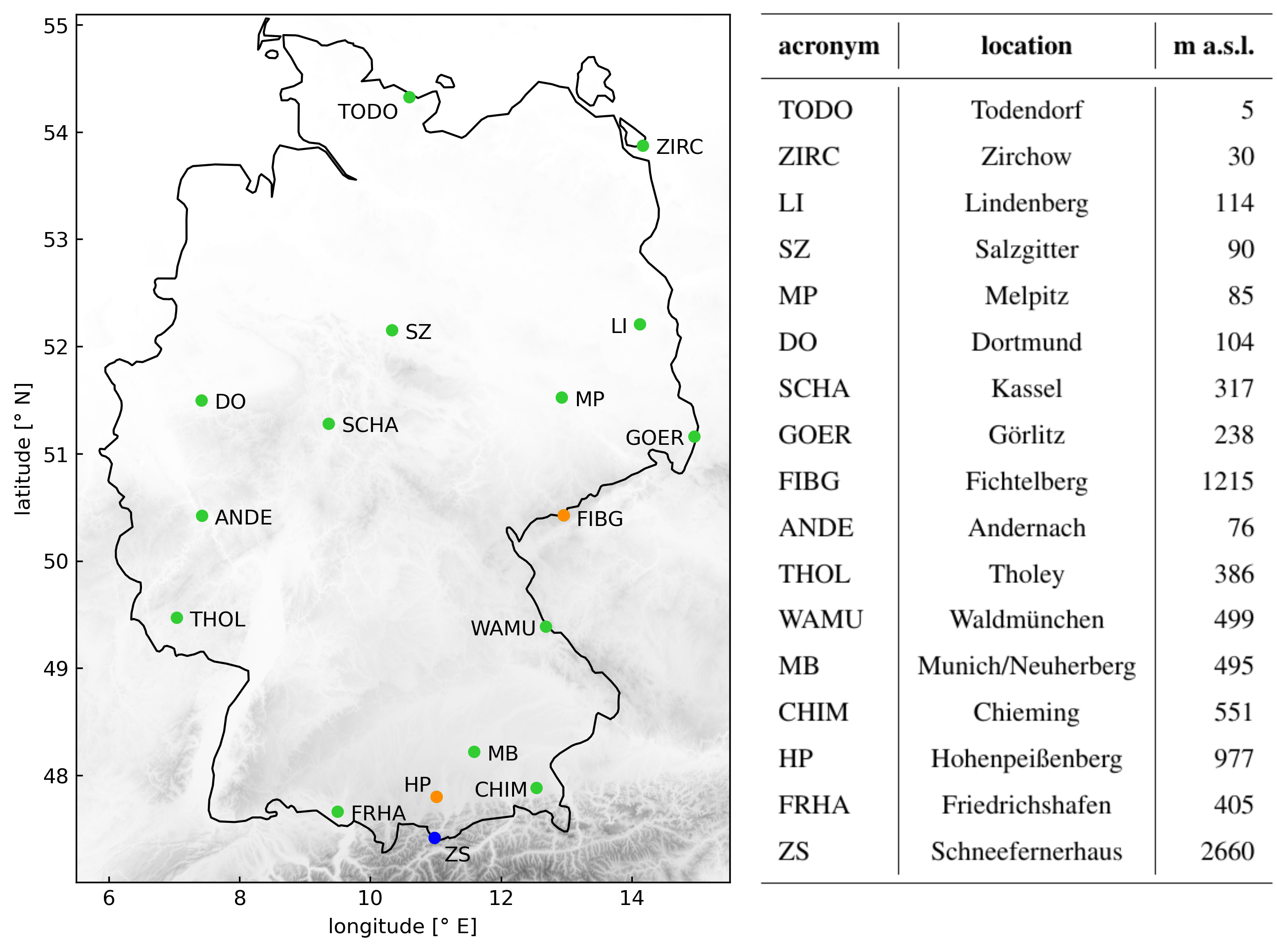}
	\caption{Locations of the measurement sites in Germany with low (green), mid (orange) and high (blue) altitude sites (left) and table of sites (right) with height above sea level and sorted from north to south}
	\label{fig: sites}
\end{figure*}

A comparison of the modelled UV Index maps with ground measurements of the German Solar Monitoring Network for the year 2022 was performed. Currently, the German Solar Monitoring Network comprises 41 measurement sites. Prior to analysis, all ground data underwent preprocessing and quality control conducted by the German Federal Office for Radiation Protection.
Finally measurements from 17 locations were selected (see Fig.~\ref{fig: sites}) to cover the area of Germany from the sea in the north to the alpine region in the south, thereby also covering different altitudes. Most of the stations (14) are located below 600~\unit{m}~a.s.l. (green markers in Fig.~\ref{fig: sites}), two of them are at a height of around 1000~\unit{m}~a.s.l. (orange markers) and the high alpine station at Schneefernerhaus is located below the summit of mount Zugspitze at 2660~\unit{m}~a.s.l. (blue marker). 

Moreover reliability of the data and coverage of the year 2022 were criteria for the selection.  
The type of measurement device used at the sites is either an sglux ERYCA broadband radiometer (site acronyms with 4 letters), a BTS2048-UV-S-WP spectroradiometer (site acronyms with 2 letters except for DO) and a Bentham DTM300 double monochromator at Dortmund (DO). 
Measurements of the UV Index were recorded every minute with the broadband devices, whereas the spectral instruments are logging a value every 2 or 6 minutes depending on the measurement site. The comparison with modelled data was then performed matching the timeline of satellite cloud information, which is available every 15 minutes, by using the ground measurements that were closest in time to the satellite recording (maximum time difference: 3 minutes). 

Since the calculation of the CMF is associated with significant uncertainties for low solar elevations, further analysis is conducted with data from solar elevations of 20° and higher. As the UV Index is small for low solar elevations (UVI < 2) anyway, the increased uncertainty in the modelled UV Index maps for this regime is insignificant.
      
\subsection{Comparison of the modelled UV Index to ground measurements}
For the following statistical evaluations, the ground measurements with a maximum time difference of 3 minutes to the satellite data acquisition were compared with the corresponding values from the modelled near-real-time UV Index map by selecting those pixels that contain the locations of the ground stations.

The greatest temporally variable influence of all atmospheric parameters on the UV Index is cloud cover. So for the following analysis the dataset was divided into values measured and calculated under clear sky conditions and those under cloudy conditions. The filtering was based on the satellite product cloud mask (CLM), where the pixel value classified the data point as cloudless or cloudy.

For the statistical comparisons, the difference between the UV Index from the calculated maps (modelled UVI) and ground measurements (measured UVI) was determined and mean and standard deviation (σ, std)  were calculated for these distributions. Moreover the correlation coefficient (corr) for measured UVI to modelled UVI was computed (see Table~\ref{tab:2}) for clear sky and cloudy sky conditions as well as for the whole (all sky) dataset.

\begin{table}[t]
	\caption{Number of data points, mean and standard deviation (std) of the differences of modelled UVI to measured UVI and correlation coefficient (corr) of measured UVI to modelled UVI for clear, cloudy and all sky conditions}
	\begin{tabular}{l|c c|r r r| c c c|c c c}
		\tophline
\textbf{site} &\multicolumn{2}{c|}{\textbf{data points}} & \multicolumn{3}{c|}{\textbf{mean}} &  
\multicolumn{3}{c|}{\textbf{std}} & \multicolumn{3}{c}{\textbf{corr}}   \\
  &\textbf{clear} &\textbf{cloud}  &\textbf{clear} &\textbf{cloud} & \textbf{all}  &\textbf{clear} &\textbf{cloud} & \textbf{all}&\textbf{clear} &\textbf{cloud} & \textbf{all}\\
\middlehline
TODO & 3208 & 5878 & -0.101 & -0.404 & -0.297 & 0.290 & 0.595 & 0.528 & 0.984 & 0.925 & 0.944 \\
ZIRC & 2978 & 6088 & 0.054 & -0.242 & -0.145 & 0.343 & 0.529 & 0.496 & 0.977 & 0.927 & 0.945 \\
LI & 2394 & 4188 & -0.374 & -0.654 & -0.552 & 0.466 & 0.703 & 0.642 & 0.976 & 0.927 & 0.944 \\
SZ & 3117 & 5999 & -0.180 & -0.535 & -0.413 & 0.364 & 0.718 & 0.643 & 0.979 & 0.918 & 0.933 \\
MP  & 3206 & 5939 & -0.284 & -0.592 & -0.484 & 0.444 & 0.685 & 0.629 & 0.972 & 0.935 & 0.942 \\
DO & 3236 & 5520 & -0.134 & -0.422 & -0.315 & 0.306 & 0.599 & 0.529 & 0.986 & 0.932 & 0.952 \\
SCHA & 1028 & 2392 & 0.080 & -0.251 & -0.151 & 0.400 & 0.638 & 0.596 & 0.974 & 0.903 & 0.929 \\
GOER & 3527 & 6005 & 0.040 & -0.318 & -0.185 & 0.389 & 0.649 & 0.593 & 0.977 & 0.907 & 0.938 \\
FIBG & 1615 & 6829 & -0.051 & -0.247 & -0.210 & 0.322 & 0.710 & 0.658 & 0.983 & 0.917 & 0.930 \\
ANDE & 3788 & 5900 & -0.179 & -0.513 & -0.383 & 0.415 & 0.732 & 0.648 & 0.981 & 0.922 & 0.944 \\
THOL & 3555 & 6206 & 0.079 & -0.262 & -0.138 & 0.352 & 0.602 & 0.550 & 0.983 & 0.941 & 0.956 \\
WAMU & 3041 & 6781 & 0.098 & -0.231 & -0.129 & 0.455 & 0.737 & 0.680 & 0.973 & 0.909 & 0.932 \\
MB & 4049 & 5705 & -0.149 & -0.441 & -0.320 & 0.364 & 0.659 & 0.574 & 0.985 & 0.929 & 0.959 \\
CHIM & 4866 & 5203 & -0.143 & -0.377 & -0.264 & 0.475 & 0.612 & 0.562 & 0.979 & 0.935 & 0.966 \\
HP & 4068 & 5704 & -0.111 & -0.323 & -0.235 & 0.411 & 0.744 & 0.636 & 0.982 & 0.915 & 0.954 \\
FRHA & 5510 & 4608 & 0.107 & -0.236 & -0.049 & 0.464 & 0.558 & 0.537 & 0.976 & 0.927 & 0.966 \\
ZS & 3540 & 6540 & -0.452 & -0.676 & -0.597 & 0.759 & 1.465 & 1.267 & 0.957 & 0.740 & 0.854 \\
\bottomhline
\end{tabular}\label{tab:2}
\end{table}

\begin{figure*}[t]  
	\includegraphics[width=12.0cm]{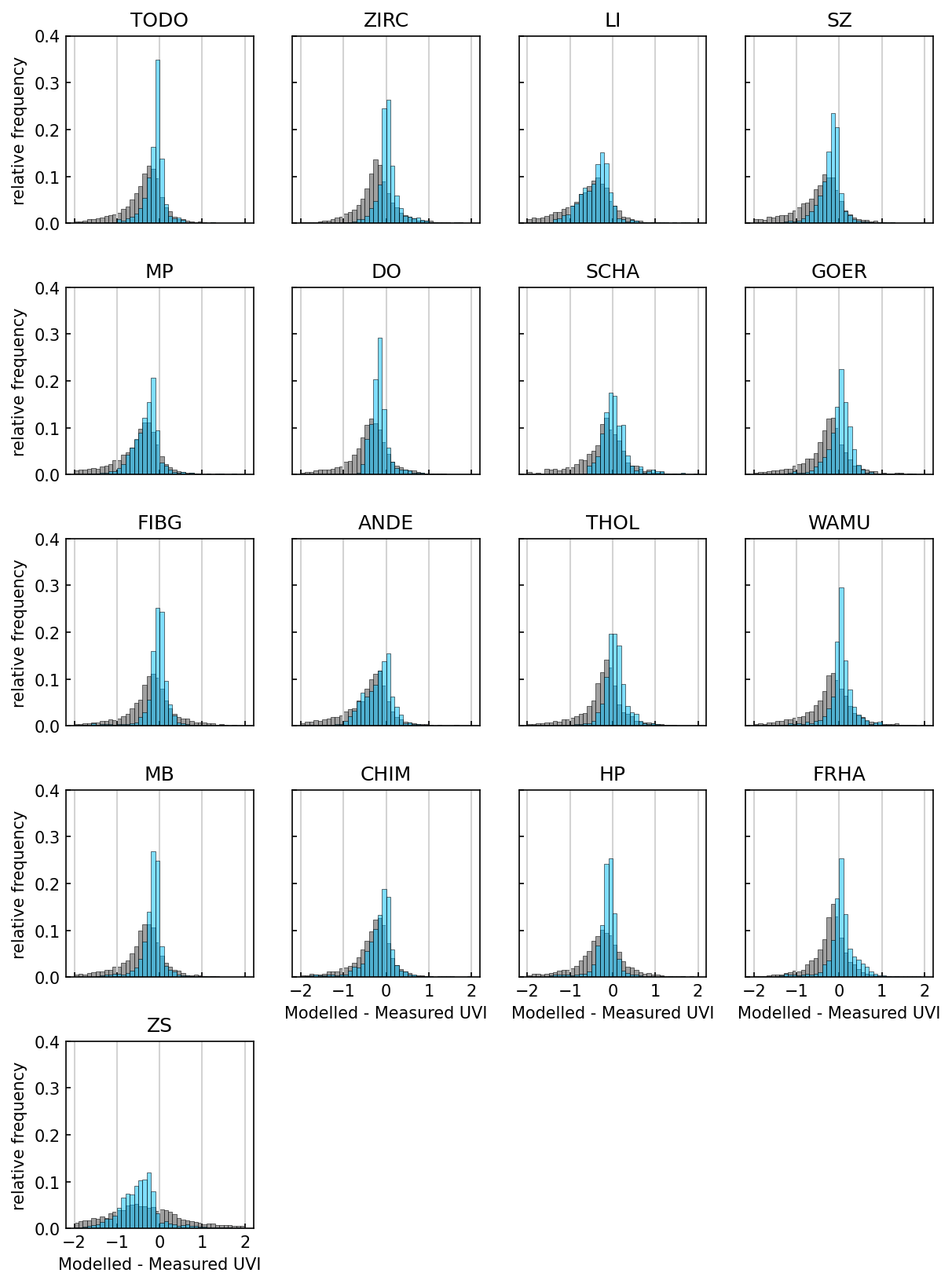}
	\caption{Histograms of the differences from modelled to measured UVI for clear sky (blue) and cloudy (grey) conditions for all ground sites}
	\label{fig: hist_clear_cloud}
\end{figure*}

The distributions of the difference modelled UVI to measured UVI for clear and cloudy sky conditions are shown in Fig.~\ref{fig: hist_clear_cloud}. Systematic differences in the histograms are clearly evident when comparing clear sky conditions (blue) and cloudy conditions (grey) for each station. For cloud cover, the standard deviation is larger, and consequently, the correlation coefficient is smaller compared to the clear sky case (see Table~\ref{tab:2}). Additionally, the mean values of the distributions for cloudy conditions are slightly shifted towards negative UVI differences, whereas for clear sky conditions they are closer to zero.      

\begin{figure*}[t]  
	\includegraphics[width=12.0cm]{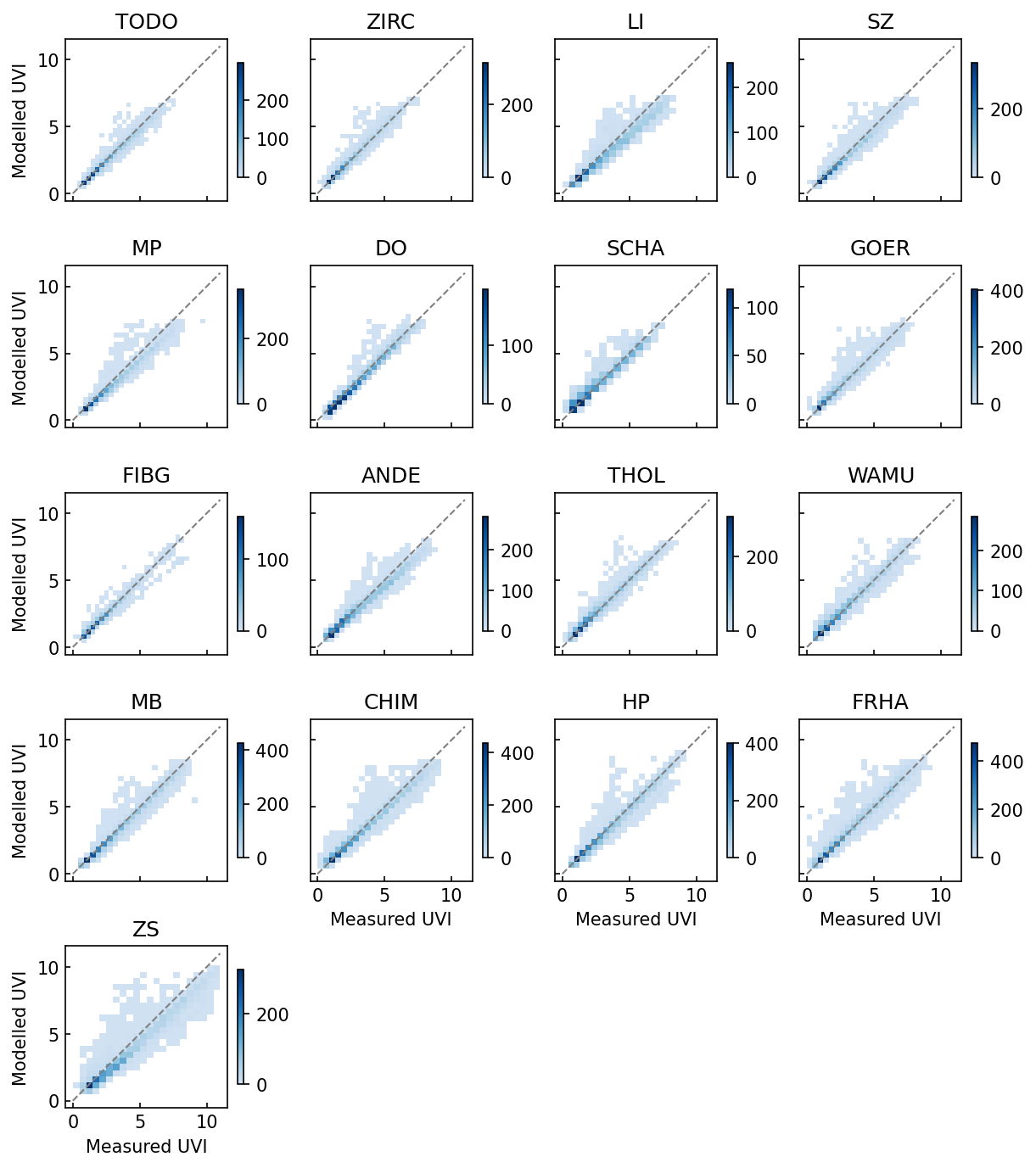}
	\caption{Correlation of measured UVI to modelled UVI for clear sky conditions for all ground sites; colours of the colour bar indicate the density of data points}
	\label{fig: corr_clear}
\end{figure*}

\begin{figure*}[t]   
	\includegraphics[width=12.0cm]{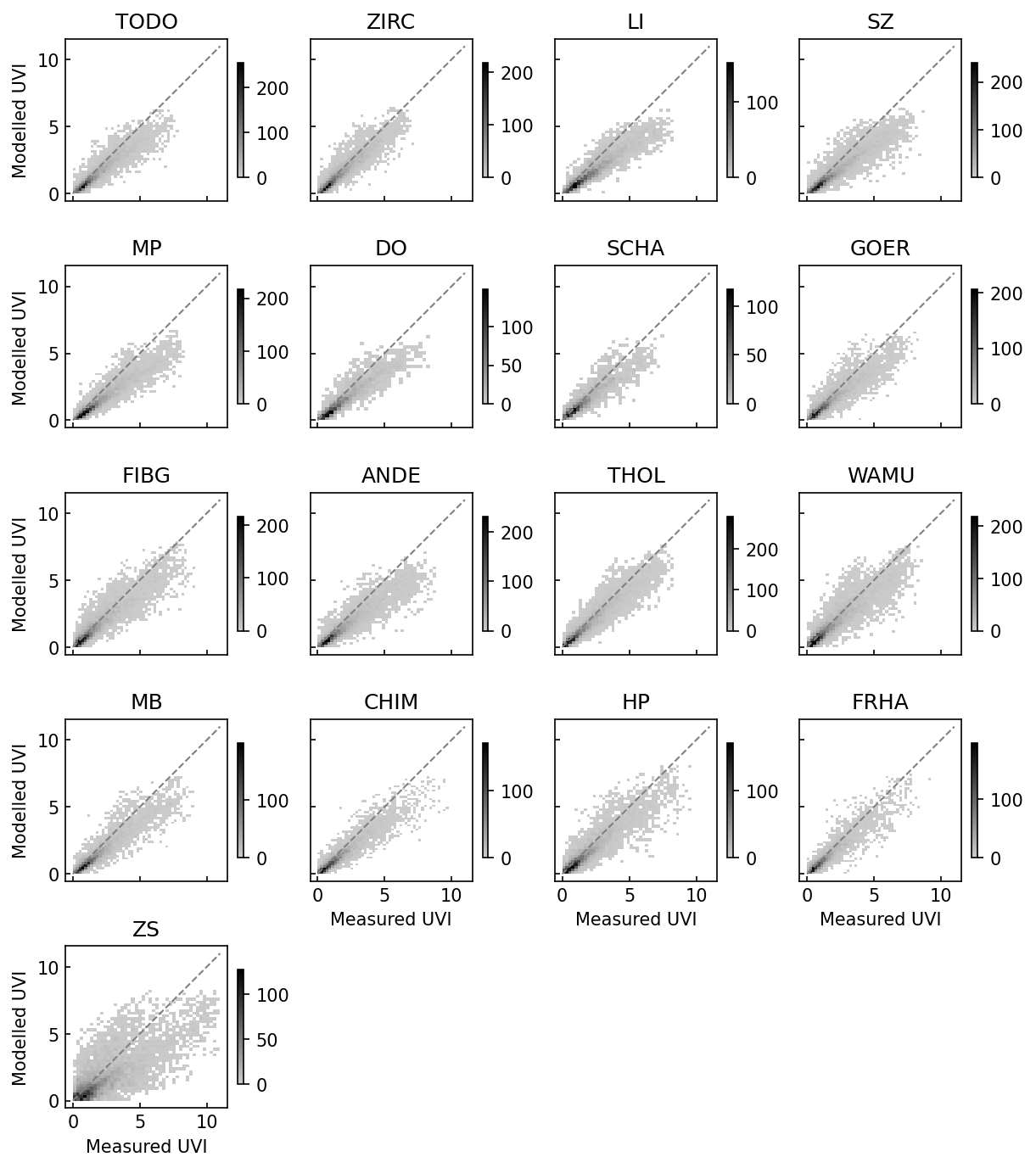}
	\caption{Correlation of measured UVI to modelled UVI for cloudy conditions for all ground sites; colours of the colour bar indicate the density of data points}
	\label{fig: corr_cloud}
\end{figure*}

The correlation of measured UVI to modelled UVI for clear sky conditions is illustrated in Fig.~\ref{fig: corr_clear} and for cloudy conditions in Fig.~\ref{fig: corr_cloud}. Naturally in both cases the density of data points is highest for low UVI following the diurnal and annual pattern of the UV Index. For all stations the spread of data points is smaller in the clear sky case compared to the cloudy one, which is also reflected in the correlation coefficients (see Table~\ref{tab:2}).

\section{Discussion}
Next we discuss the peculiarities of the results in Table~\ref{tab:2} with regards to clear sky and cloudy conditions and the impact of topography. Then the results are compared with the recently published UVIOS model (\cite{Kosmopoulos21}). Finally we discuss potentially beneficial combinations of ground-based measurements and modelled UV Index maps.  

\subsection{Clear sky conditions}
The overall agreement of the clear sky model with the ground measurements is quantified in terms of the mean and standard deviation of the distribution of UVI differences (modelled UVI - measured UVI) (see Table~\ref{tab:2}). Considering the mean values in the clear sky case, which range from -0.452 (ZS) to 0.107 (FRHA), no overall bias of the model is detected.
These deviations of the mean value from zero are attributed to calibration uncertainties of the ground measurements as well as to pollution of the measurement devices. Moreover, uncertainties in the clear sky model are introduced by using forecast data from CAMS (ozone, albedo, Angstrom $\beta$) with a relatively coarse resolution of 0.4° (approximately 28~\unit{km} x 45~\unit{km}) as input for the LUT interpolation. 
Especially in mountainous terrain with a highly variable topography the rough granularity of model input data is leading to higher uncertainties. This effect becomes apparent, when comparing the statistical parameters of the high mountain station Schneefernerhaus (ZS)
with the results of other sites (see Fig.~\ref{fig: hist_clear_cloud} and~\ref{fig: corr_clear}). In each case, the other stations exhibit more favourable statistics than ZS (see Table~\ref{tab:2}).
 
\subsection{Cloudy conditions}\label{sec: clouds}
Especially with temporally variable cloud cover, it becomes apparent that due to the relatively coarse temporal (15 minutes) and spatial resolution (0.06°, 4~\unit{km}~x~7~\unit{km}) of cloud information used for CMF calculation and cloud detection, there are larger discrepancies between ground measurements and modelled UV Index for cloudy conditions. This results in larger standard deviations in the difference between modelled and measured UVI (see Fig.~\ref{fig: hist_clear_cloud}). On average, the model is underestimating the UVI in the presence of clouds for all stations in the range of -0.231 UVI (WAMU) up to -0.676 UVI (SZ) and also the correlation coefficients are decreasing (see Table~\ref{tab:2}) compared to the cloudless case. 

\begin{figure*}[t]  
	\includegraphics[width=12.0cm]{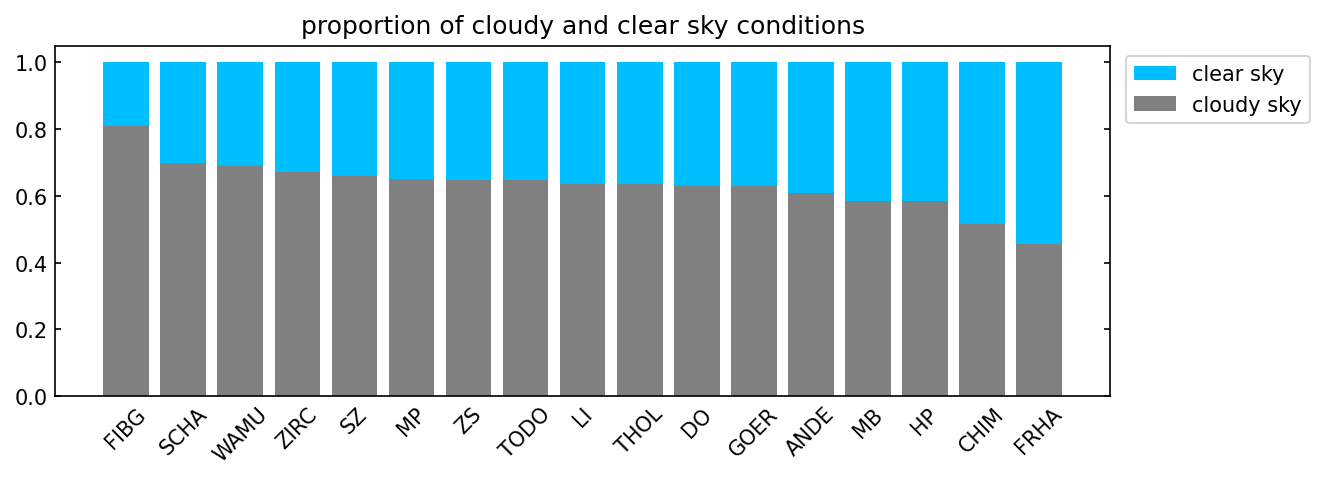}
	\caption{Proportion of measurements with cloudy (grey) and clear sky (blue) conditions for each ground station for the year 2022}
	\label{fig: clear_cloud}
\end{figure*}

When assessing the concordance between modelled and measured UV Index values across all sky conditions (both clear and cloudy), the uncertainty is considerably influenced by the climatology of cloudiness. Hence, Fig.~\ref{fig: clear_cloud} illustrates the proportion of measurements with and without cloud cover at each station throughout the year 2022. Across the majority of sites, the proportion of cloudy measurements falls within the range of 60\% to 65\%. Consequently, for these stations, the impact of cloudiness on the uncertainty of the whole dataset (see column 'all' in Table~\ref{tab:2}) is quite similar. 
At Fichtelberg (FIBG), approximately 80\% of the measurements were classified as cloudy, resulting in a standard deviation $\sigma = 0.658$ of the all sky UVI difference that is close to the corresponding value for cloudy sky scenarios ($\sigma = 0.710$). In contrast, the standard deviation for the clear sky UVI difference in FIBG is significantly smaller with $\sigma = 0.322$.
Friedrichshafen (FRHA) is the only site with a majority of clear sky measurements (5510 compared to 4608 cloudy sky measurements), but this prevalence (55\% versus 45\%) is too small to have a significant impact on the statistics of the all sky distribution.

Measured data from the ground stations Munich/Neuherberg (MB) and Hohenpeißenberg (HP) are suitable for comparisons due to their relatively close proximity of approximately 50~\unit{km} linear distance. Nevertheless, for the Munich/Neuherberg station situated at the northern outskirts of the city, higher Aerosol Optical Depth (AOD) levels are anticipated in comparison to Hohenpeißenberg, which is located in a rural area south-west of Munich. Moreover, the measurement station at Hohenpeißenberg is located approximately 500~\unit{m} higher than the station in Munich/Neuherberg. So the ground measurements of MB are expected to be a bit lower than those from HP for clear sky conditions.
In Fig.~\ref{fig: HP,MB clouds}, measured and modelled UVI data from both stations are compared on days with partly cloud cover. The impact of clouds on ground measurements is clearly visible in the diurnal variation of the UVI (top row). On May 31, 2022 both stations were affected by clouds approaching from the east (since MB is affected by cloudiness earlier in the day than HP) and on June 4, 2022, Hohenpeißenberg was less affected by cloud cover in the forenoon than the station in Munich/Neuherberg. This is also evident in the respective middle graphs, where green and blue dots are representing times with cloud cover according to the cloud mask.
The ratios of measurements at MB to HP and modelled UVI at MB to HP do match quite well (red and orange lines), with only small deviations due to higher uncertainties in the model calculations under cloudy conditions.
The UVI difference of model to measurement for both sites is depicted in the bottom row and shows a good agreement especially for clear sky conditions of about 0.6 UVI and even for cloudy conditions most of the UVI differences are within 1~UVI. 

\begin{figure}[t]
\includegraphics[width=12.0cm]{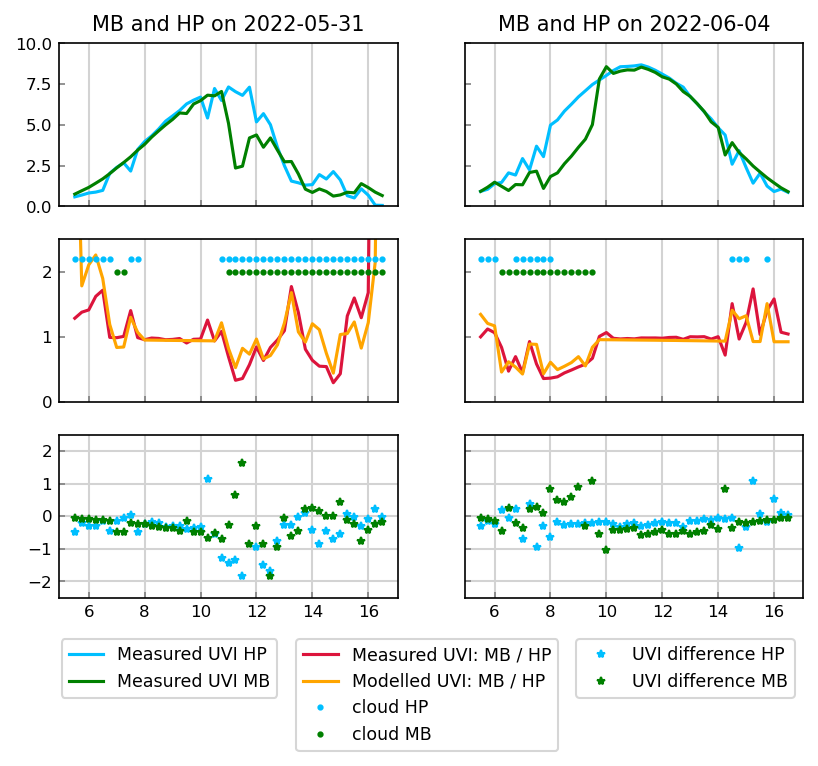} 
\caption{Comparison of MB and HP on days with partly clouds: measured UV Index from MB and HP (top row, left legend), ratio of measurements MB / HP with cloud filtering for both sites (middle row, middle legend), difference of modelled UVI and measured UVI for both sites (bottom row, right legend)}
\label{fig: HP,MB clouds}
\end{figure}

\begin{figure*}[t]
  \includegraphics[width=12cm]{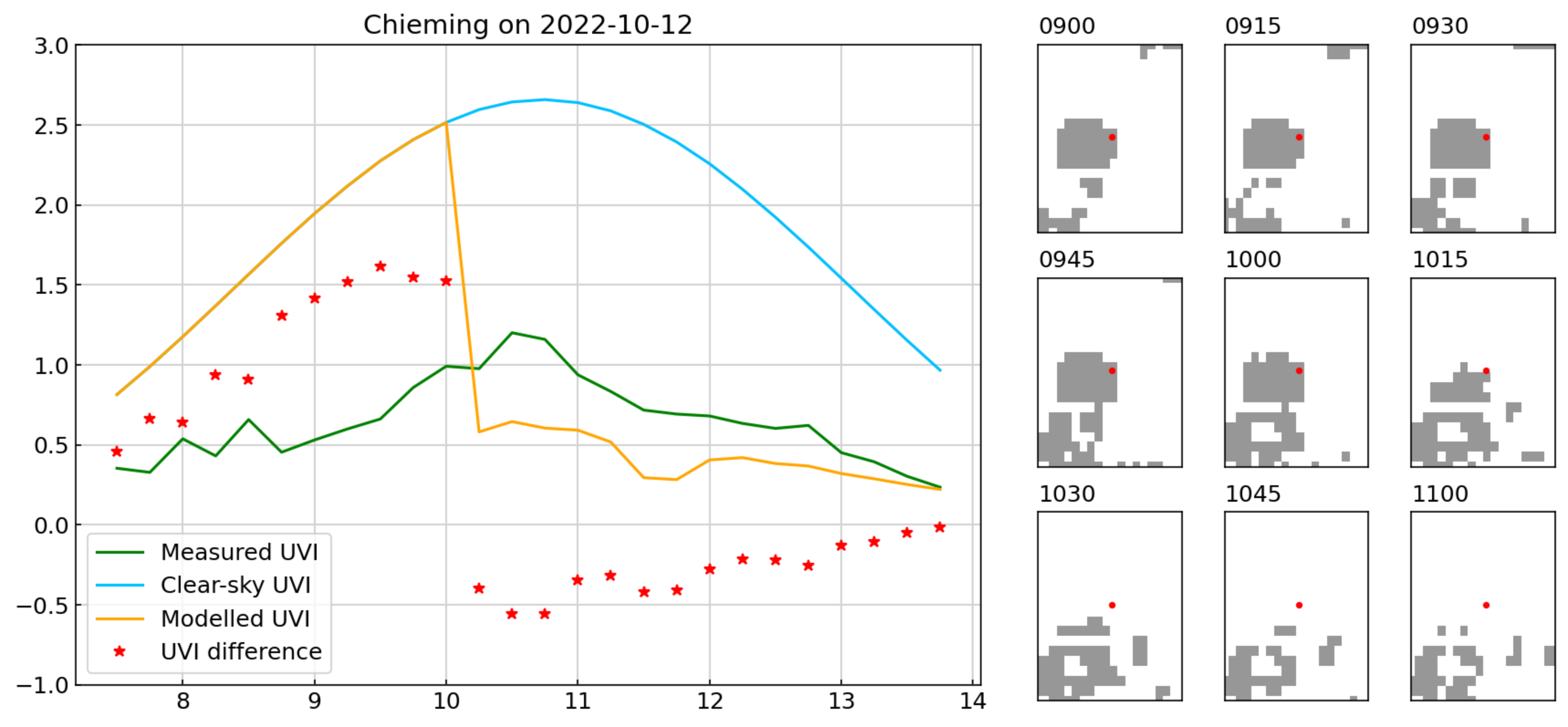}
  \caption{Comparison of measured (green) and modelled (orange) UVI at the site Chieming (CHIM) for day 2022-10-12 (left) and a selection of CLM (cloud mask) details (right) to illustrate the cloud situation in the vicinity of the site}
  \label{fig: CHIM}
  \end{figure*}

In Fig.~\ref{fig: CHIM} a day (2022-10-12) with clouds in the vicinity of site Chieming (CHIM) is illustrated to show the limitations of the model due to the filtering in clear and cloudy situations via the satellite product cloud mask. When comparing the measured UVI (green) to the clear sky model (blue) it is obvious that the  ground measurements were affected by clouds during the whole day. Also the cloud mask is indicating clouds (white areas in the right timeline) to the east and the north of the site, nevertheless the pixel corresponding to the site location (red dot) is showing cloudless conditions up to 10:00~UTC. So for the model calculation cloudless conditions are assumed until 10:00~UTC, leading to a vast overestimation of the model (orange) compared to the actual measurements (green) and therefore to an increased UVI difference (red stars) in the range of 0.5~UVI up to 1.7~UVI. After 10:00~UTC the cloud mask is indicating cloud cover also for the pixel of CHIM, thus the application of the CMF is decreasing the modelled UV Index resulting in a smaller UVI difference of about 0.5~UVI.
In this case the sun, positioned to the east of the site in the forenoon, was obstructed by clouds in neighbouring pixels. Due to the filtering of the site pixel as unclouded, these data are increasing the standard deviation of the cloudless distribution of the UVI differences.

Of course, misleading filtering also impairs the statistics of the distribution for cloudy conditions, e.g. 
for broken cloud situations, where the CLM is indicating a cloud for the pixel of the site while the direct sun may not be affected by clouds at all. For such conditions even an enhancement of the measured UVI compared to the clear sky expectation is possible (\cite{PfisterEA03}, \cite{CalboEA05}).      

\subsection{Impact of the topography}
Most of the stations of the German Solar Monitoring Network used for this study (see Fig.~\ref{fig: sites}) are located below 600~\unit{m}~a.s.l. (see Fig.~\ref{fig: sites}). Two sites, Hohenpeißenberg (HP, 977~\unit{m}~a.s.l.) and Fichtelberg (FIBG, 1215~\unit{m}~a.s.l.), are situated in the low mountain range and the high alpine station Schneefernerhaus (ZS) is positioned at 2660~\unit{m}~a.s.l..
When comparing the statistical parameters of HP and FIBG in Table~\ref{tab:2} to the values of low altitude sites, no significant variation can be found, however this is not true for the site Schneefernerhaus.
The distributions of the UVI differences from the site Schneefernerhaus (ZS) show the largest standard deviations and biggest shifts of the mean value from zero (see Table~\ref{tab:2}) compared to the results of all other stations for clear sky as well as cloudy conditions and consequently also for the analysis of the all sky dataset. Accordingly, the correlation coefficients are smaller for site ZS, not only but especially for cloudy conditions with 0.740 compared to the correlation coefficients of the other stations, all of which are above 0.9. 

This is attributed to the pronounced topography in the mountains, which is characterized by spatially small-scale variable albedo and cloud situations. Under such conditions, the predicted albedo data from CAMS, which provides an average value over an area of about 28~\unit{km} x 45~\unit{km} (pixel resolution 0.4°), and the cloud information  derived from satellite images, available at a resolution of approximately 4~\unit{km} x 7~\unit{km} (0.06°), are not representative for the actual conditions at the site.

\subsection{Comparison of the results with the UVIOS model}
A similar approach for calculating UV Index maps for Europe was presented in \citet{Kosmopoulos21}, where satellite data from MSG was utilized for cloud analysis, combined with ozone, aerosol, and ground albedo data as input parameters for a solar irradiance lookup table to compute near-real-time UVI maps. The so called UVIOS (UV Index Operating System) model uses forecasted and measured atmospherical properties from eight different sources in combination with high performance computing architectures to speed up the retrieval of UV Index maps with 1.5~million pixels covering Europe. The calculated UV Index values from UVIOS were compared with measurements from 17 European ground stations for the year 2017. Also the most temporally proximate ground measurements, with a maximum time difference of 3 minutes, were employed. The publication describes the percentages of UVI differences (UVIOS minus measured UVI) within the ranges ±0.5~UV~Index (U0.5) and ±1.0~UV~Index (U1.0) in Table~3 of \citet{Kosmopoulos21}. It also provides the correlation coefficient for cloudless conditions and for the entire dataset (both starting at a solar elevation angle of 20°).

To conduct a comparison, the same parameters for the differences of modelled UVI to measured UVI for the stations of the German Solar Monitoring Network can be found in Table~\ref{tab:3}. Since the accuracy of both models depends on the statistics of cloud cover in the respective years (UVIOS ground measurements from 2017, German Solar Monitoring Network from 2022), individual statistics cannot be compared directly. Therefore averages of U0.5 and U1.0 were computed for both models. In the clear sky case the UVIOS results for U0.5 (88,5\%) and U1.0 (95,67\%) are very similar to the results of this study (see Table~\ref{tab:3}). This is also the case for the mean correlation coefficient of 0.95 (UVIOS) and 0.98 (this study) for clear sky conditions as well as for all sky conditions (0.93 for UVIOS and 0.94 for this study).
However, the mean values of U0.5 and U1.0 in the all sky case of this study are lower than the UVIOS results: UVIOS U0.5 being 82,85\% versus 69,13\% and for U1.0 93,26\% versus 87.71\%. As mentioned above (see Section~\ref{sec: clouds}) the impact of cloud cover on the statistics of the all sky case is dependent on the amount and the mainly occurring patterns of cloudiness in the respective years (2017 and 2022). Thus, in the all sky case, the comparison of the mean values of U0.5 and U1.0 for both studies has little informative value.

\begin{table}[t]
	\caption{U0.5 and U1.0 of the difference of modelled UVI to measured UVI and average of U0.5 and U1.0 across all stations}
	\begin{tabular}{l|c c|c c}
		\tophline
		\textbf{site} &\multicolumn{2}{c|}{\textbf{clear}} & \multicolumn{2}{c}{\textbf{all}} \\
		&\textbf{U0.5} &\textbf{U1.0} &\textbf{U0.5} &\textbf{U1.0} \\
		\middlehline
	TODO & 73.79 & 90.72 & 91.24 & 98.66 \\
	ZIRC & 77.52 & 93.86 & 91.03 & 98.39 \\
	LI & 52.54 & 81.91 & 61.99 & 91.81 \\
	SZ & 64.92 & 85.76 & 85.79 & 96.79 \\
	MP & 60.48 & 85.10 & 72.40 & 95.13 \\
	DO & 69.01 & 89.37 & 85.58 & 97.86 \\
	SCHA & 74.21 & 89.94 & 89.20 & 96.21 \\
	GOER & 75.50 & 90.12 & 90.50 & 96.60 \\
	FIBG & 71.17 & 88.00 & 93.00 & 96.78 \\
	ANDE & 65.72 & 87.26 & 76.35 & 96.86 \\
	THOL & 78.68 & 91.93 & 91.20 & 98.23 \\
	WAMU & 71.84 & 88.05 & 86.58 & 95.23 \\
	MB & 73.21 & 89.42 & 89.28 & 96.37 \\
	CHIM & 75.53 & 90.71 & 83.11 & 93.94 \\
	HP & 74.27 & 89.36 & 90.34 & 95.38 \\
	FRHA & 79.52 & 93.20 & 82.65 & 94.57 \\
	ZS & 37.23 & 66.41 & 46.95 & 83.42 \\
		\middlehline
	\textbf{average} & 82.78 & 95.42 & 69.13 & 87.71 \\	
		\bottomhline
	\end{tabular}\label{tab:3}
\end{table}

\subsection{Combination of ground-based measurements and modelled UV Index maps}

There are beneficial prospects in integrating ground-based measurements, which provide real-time information about the current UV radiation levels in the vicinity of a site, with modelled UV Index maps based on satellite imagery. While ground-based measurements offer continuous data, modelled UV Index maps provide extensive coverage across larger areas but only at specific time intervals.

First modelled data for cloudless sky conditions can be used for quality control of the ground-based measurements. When comparing the ratio of measured UVI to the corresponding pixel of the modelled UV Index maps, periods with larger deviations can help to identify problems with the ground-based measurement devices. In particular, if the deviations only occur in the data of one station, this is a strong indication of possible contamination of the detectors or some form of defect. 

Furthermore, modelled UV Index maps can provide climatological data, particularly for regions that are not in close proximity to a ground station. Maintaining a high density of ground stations is costly and time consuming. Receiving near-real-time measurements from SEVIRI / MSG also involves additional infrastructure and processing, but the effort is much lower compared to maintaining a ground-based network.

By using geospatial interpolation techniques, area covering information can be gained from ground-based measurements (\cite{Schmalwieser2001}, \cite{Duguay1995}). Nevertheless, without area-wide cloud information these UV Index maps are associated with large uncertainties. So the goal is to combine both datasets to benefit on the one hand from the accuracy of measured UV levels at the ground and on the other hand from the area-wide coverage provided by satellite-based UV Index maps.

To explore the possibilities of a spatial correction of the modelled UV Index maps through ground measurements, days with very low cloud coverage in the area of Germany were selected. Next for each point in time (15 minute intervals) the subset of clear sky data from all 17 sites was filtered via the cloud mask. Then the ratio of these data points to the corresponding pixel values of the UV Index maps was analysed. While most of the data points lie within 0.8 to 1.2, there are some outliers that are
attributed to uncertainties in the model calculation due to forecasted and spatially coarse input parameters or to faulty cloud filtering (compare Fig.~\ref{fig: CHIM}). So the ratio data points were limited to the range from 0.8 up to 1.2 and then geostatistically interpolated to obtain correction maps for the UV Index maps. The Kriging interpolation method (\cite{Oliver1990}, \cite{Meng2013}) was initially evaluated, for which the spatial correlation of the data points is crucial. However, the correlation of a maximum of 17 clear sky data points, widely distributed in the area of Germany, proved insufficient to yield meaningful results with this method.

Therefore, the RBFInterpolator method (Radial Basis Function Interpolator) from the Python SciPy package was then used for calculating the spatial distribution. This method is suitable for interpolating multidimensional data points. The method is based on radial basis functions (RBF) with a linear kernel, whose value depends only on the distance from the centre of the function. A deliberately simple approach with a plausible distance weighting was chosen for the calculation of the spatial distribution, as it will be shown in the following that the fundamental issue does not lie in the details of the spatial interpolation method.

\begin{figure*}[t]
	\includegraphics[width=12cm]{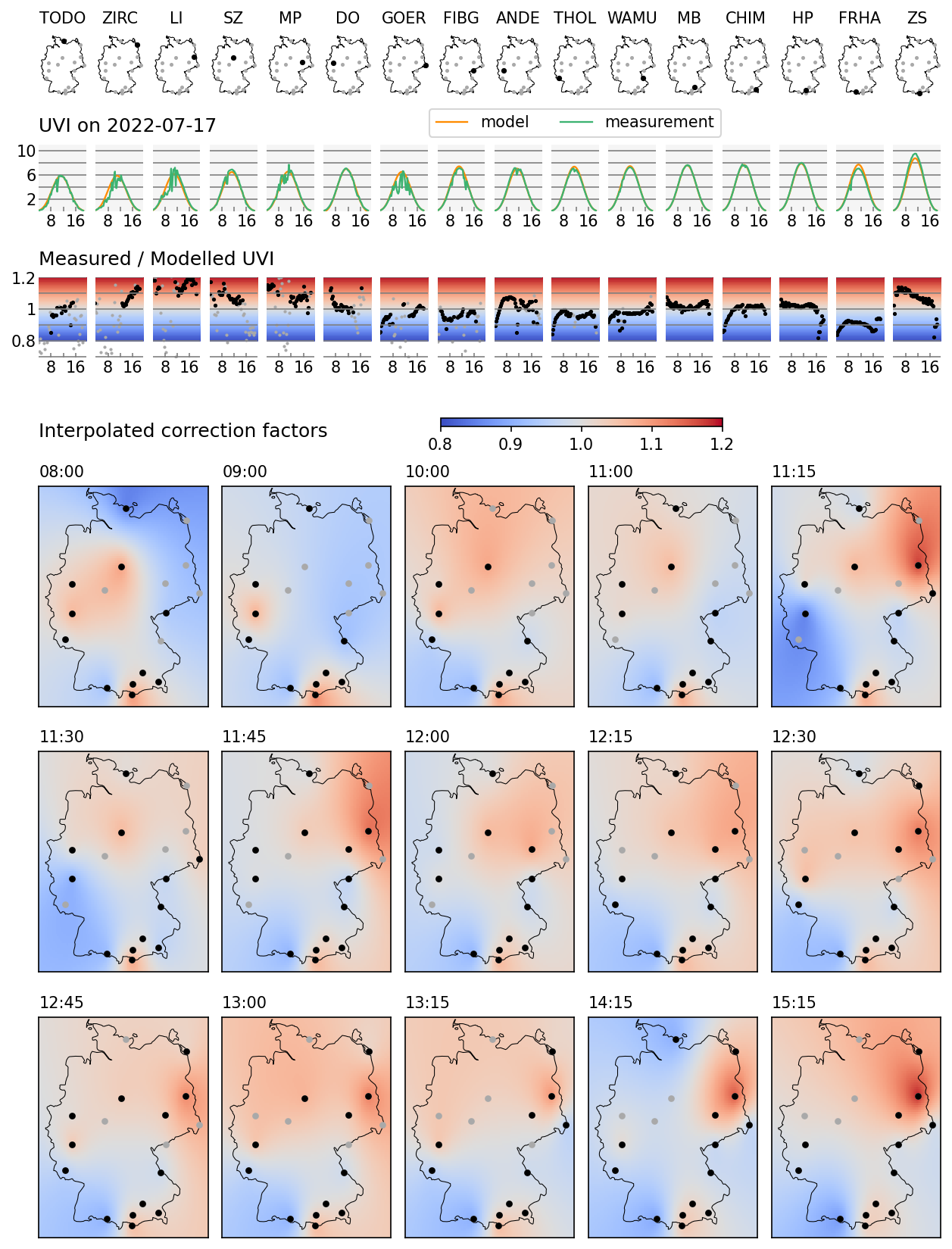}
	\caption{From top to bottom: Location of the ground station, daily course of measured and modelled UV Index, ratio of measurement to model, and spatially interpolated correction factors for July 17, 2022}
	\label{fig: corr_map}
\end{figure*}

The results for July 17, 2022 are illustrated in Fig.~\ref{fig: corr_map}, showing the course of measured and modelled UV Index values per station, along with the resulting ratios (upper section). In the representation of the ratios, the data points for cloudless conditions are depicted in black. These values form the basis for the calculation of correction maps (lower part of the figure) through spatial interpolation at individual points in time.

On this day, spatial interpolation can be performed using measured values from 16 ground stations throughout the day, although never all stations were unclouded simultaneously (see black points in the interpolated correction factor maps). Considering the timeline of correction factor maps (lower part of Fig.~\ref{fig: corr_map}), the spatial distribution is varying strongly with time, showing the effect of missing data points (grey dots). For example at 9:00 UTC the station Salzgitter (SZ) is obstructed by clouds and therefore missing in the calculation of the correction map, leading to a correction factor of slightly below 1 in the northern part of Germany, while before and after that the correction is bigger than 1 in that area.   
The data from Todendorf (TODO) are also used only intermittently, resulting in an inhomogeneous temporal pattern of correction maps.
In comparison to the model, the data from Friedrichshafen (FRHA) are consistently lower at approximately 0.9 but remain constant throughout the day. The closely located stations Munich/Neuherberg (MB), Chieming (CHIM) and Hohenpeißenberg (HP) all show a similar correction factor that remains constant throughout the day, which increases the confidence in the correction factors in the vicinity of these stations.

In essence, it is evident that each ground station has the potential to exert a significant influence on the correction maps, and a variable number of used data points for interpolation results in timely variable corrections. This raises the question of how and at what frequency corrections should judiciously be applied. Uncertainties arise both in ground measurements, which typically undergo post hoc verification and potential correction, and in the input parameters for the model calculation, initially provided as forecast values and only later available in higher quality through measurements.

When considering the combination of ground-based measurements with modelled UV Index maps for cloudy conditions the shortcomings become even worse. When the intra pixel variability of clouds becomes significant, ground-based data can vary from very low UV Index to values even higher than the clear sky model expectations. However, the modelled UV Index, which is relying on the information of a pixel to be cloudy, will always be smaller than the clear sky model expectation. This circumstance is the main factor leading to the enlarged standard deviations of the UVI difference for cloudy conditions compared to the clear sky case (see Table~\ref{tab:2}). In principle for broken cloud conditions satellite derived UV Index maps lack the representativeness for ground-based measurements and vice versa.

\conclusions  
In this study, the method to calculate near-real-time UV Index maps (\cite{amt-2023-188}), that is used operationally in the framework of the Austrian UV Monitoring Network, is employed to calculate UV Index maps for the area of Germany. Clear sky UV Index maps are retrieved from a pre-calculated lookup table of erythemally weighted UV radiation by employing forecasted data of total ozone column, albedo and aerosol optical depths from the CAMS global model forecast. To incorporate the effects of clouds in the model, a cloud modification factor (CMF) is derived from SEVIRI imagery, that is recorded every 15 minutes onboard the MSG satellite. 

With this method, almost 22.000 UV Index maps of Germany have been calculated for the year 2022. Ground-based measurements of 17 stations of the German Solar Monitoring Network where then compared to the corresponding pixel value of the modelled maps. Statistical analysis of the UVI differences (modelled minus measured UVI) show a good agreement between model and measurements for clear sky conditions with almost 83\% of all clear sky UVI differences being within 0.5~UVI and 95\% within 1.0~UVI (see Table~\ref{tab:3}). Also the correlation between measured and modelled UVI is higher than 0.95 for all sites (see Table~\ref{tab:2}).

For cloudy conditions the statistics worsen, with only 69\% of the UVI differences being within 0.5~UVI and 87\% within 1.0~UVI (see Table~\ref{tab:3}). This is due to the uncertainties arising from the cloud information retrieval from satellite data, which can not properly represent the actual cloud situation affecting the solar radiation at the location of the ground site.
Especially for broken cloud conditions, the pixel size of 4~\unit{km} x 7~\unit{km} (0.06° resolution) of the satellite data is insufficient to reproduce the effects of small scale clouds. 
 
Not only clouds but also highly variable topography (e.g. in mountainous regions) is adding to the uncertainties in the modelled UV Index maps. The statistics of the site Schneefernerhaus (ZS), located at 2660~\unit{m}~a.s.l. in the alpine region, are worse than the corresponding values of all the other sites and for all considered conditions (clear, cloudy and all sky). With a highly variable small-scale topography, albedo predictions from CAMS with a pixel size of approximately 28~\unit{km} x 45~\unit{km} (0.4° resolution) can not represent the actual albedo situation influencing measurements at the location of the high alpine site.

So, for cloudy conditions (especially broken clouds) and small-scale variable topography satellite derived UV Index maps lack the representativeness for ground-based measurements and vice-versa. Nevertheless, there are beneficial prospects in combining ground-based measurements and modelled UV Index maps. Ground-based measurements provide continuous UV radiation levels at the ground, but restricted in their significance to the vicinity of a site, whereas 
modelled UV Index maps based on satellite imagery offer extensive area coverage but only at specific time intervals.

The possibility of integrating both datasets via a correction factor map based on the ratio of modelled UVI to measured UVI was tested for clear sky conditions. By spatially interpolating the ratio of clear sky modelled to measured UVI  for each point in time (15 minute interval), correction factor maps with values from 0.8 to 1.2 were generated for days with low cloud coverage. The case study presented for July 17, 2022 shows that the spatial distribution of the correction factor maps is varying strongly with time because the geostatistical interpolation is sensitive to the number of supporting points. With a loose cloud field passing through, the data of a site are only used intermittently, leading to an inhomogeneous correction factor throughout the day. Moreover, each ground station has the potential to exert a significant influence on the correction maps (see Fig.~\ref{fig: corr_map}). If considering a correction for cloudy conditions all the above mentioned discrepancies add to the problem. 
This leaves us with the question of how and at what frequency corrections should judiciously be applied. Naturally, the answer to this question depends on the specific application. So, for the improvement of near-real-time UV Index maps a different approach for the correction may be expedient than for analysing the data in retrospect.  
Uncertainties emerge in both ground measurements, commonly subjected to post hoc validation and potential adjustments, and in the input parameters for model calculations, initially supplied as forecast values and later refined to higher quality through subsequent measurements.
 
To further explore possible synergies of ground-based measurements and satellite derived maps more investigations on the proper methods to combine them are planned.

\authorcontribution{BK processed the UV index measurements of the German Solar Monitoring Network, compared them with model results, derived the correction factor maps and was responsible for quality assurance. MS calculated the clear-sky lookup table. VS designed and implemented the CMF algorithm. The manuscript and figures were drafted by BK, with all authors contributing to its final state.} 

\competinginterests{The authors declare no competing interests.}

\begin{acknowledgements}
The authors thank Sebastian Lorenz of the German Federal Office for Radiation Protection (BfS) for providing most of the ground data of the German Solar Monitoring Network (including preprocessing and quality control of all ground data used in this study) and for valuable discussions. \\
The authors thank the Meteorological Observatories Lindenberg and Hohenpeißenberg of the German National Meteorological Service (DWD) for providing ground measurements of the stations Lindenberg and Hohenpeißenberg, the Federal Institute for Occupational Safety and Health (BAuA) for providing ground measurements of the station Dortmund and the Leibniz Institute for Tropospheric Research (TROPOS) for providing ground measurements of the station Melpitz. \\
The authors thank the German Federal Office for Radiation Protection (BfS) and the German Federal Ministry for the Environment, Nature Conservation, Nuclear Safety and Consumer Protection (BMUV) for supporting this study. \\
The Austrian UV network is supported by a grant from the Federal Ministry for Climate Action, Environment, Energy, Mobility, Innovation and Technology (BMK). \\
This publication uses Copernicus Atmosphere Monitoring Service Information [2023]; neither the European Commission nor ECMWF is responsible for any use that may be made of the information it contains.
\end{acknowledgements}

\bibliographystyle{copernicus}
\bibliography{ref.bib}

\begin{thebibliography}{21}
\providecommand{\natexlab}[1]{#1}
\providecommand{\url}[1]{\texttt{#1}}
\providecommand{\urlprefix}{}
\expandafter\ifx\csname urlstyle\endcsname\relax
  \providecommand{\doi}[1]{https://doi.org/\discretionary{}{}{}#1}\else
  \providecommand{\doi}{https://doi.org/\discretionary{}{}{}\begingroup
  \urlstyle{rm}\Url}\fi

\bibitem[{Calbó et~al.(2005)Calbó, Pagès, and González}]{CalboEA05}
Calbó, J., Pagès, D., and González, J.-A.: Empirical studies of cloud
  effects on UV radiation: A review, Reviews of Geophysics, 43,
  \doi{10.1029/2004RG000155}, 2005.

\bibitem[{CAMS(2021)}]{CAMSdata}
CAMS: CAMS global atmospheric composition forecasts,
  \url{https://ads.atmosphere.copernicus.eu/cdsapp#!/dataset/cams-global-atmospheric-composition-forecasts?tab=overview},
  last accessed 2022/12/12, 2021.

\bibitem[{Duguay(1995)}]{Duguay1995}
Duguay, C.~R.: An approach to the estimation of surface net radiation in
  mountain areas using remote sensing and digital terrain data, Theoretical and
  Applied Climatology, 52, 55--68, \doi{10.1007/BF00865507}, 1995.

\bibitem[{Emde et~al.(2016)Emde, Buras-Schnell, Kylling, Mayer, Gasteiger,
  Hamann, Kylling, Richter, Pause, Dowling, and Bugliaro}]{libradtran2}
Emde, C., Buras-Schnell, R., Kylling, A., Mayer, B., Gasteiger, J., Hamann, U.,
  Kylling, J., Richter, B., Pause, C., Dowling, T., and Bugliaro, L.: The
  libRadtran software package for radiative transfer calculations (version
  2.0.1), Geoscientific Model Development, 9, 1647--1672,
  \doi{10.5194/gmd-9-1647-2016}, 2016.

\bibitem[{EUMETSAT(2015)}]{clm}
EUMETSAT: MSG Meteorological Products Extraction Facility Algorithm
  Specification Document, EUMETSAT, doc.No. EUM/MSG/SPE/022, Issue v7B
  e-signed, 2015.

\bibitem[{EUMETSAT(2017)}]{seviri}
EUMETSAT: MSG Level 1.5 Image Data Format Description, EUMETSAT, doc.No.
  EUM/MSG/ICD/105, Issue v8 e-signed, 2017.

\bibitem[{Gesch et~al.(1999)Gesch, Verdin, and Greenlee}]{GeschEA99}
Gesch, D.~B., Verdin, K.~L., and Greenlee, S.~K.: New land surface digital
  elevation model covers the Earth, Eos, Transactions American Geophysical
  Union, 80, 69--70, \doi{10.1029/99EO00050}, 1999.

\bibitem[{Juzeniene et~al.(2011)Juzeniene, Brekke, Dahlback, Andersson-Engels,
  Reichrath, Moan, Holick, Grant, and Moan}]{Juzeniene2011}
Juzeniene, A., Brekke, P., Dahlback, A., Andersson-Engels, S., Reichrath, J.,
  Moan, K., Holick, M.~F., Grant, W.~B., and Moan, J.: Solar radiation and
  human health, Reports on Progress in Physics, 74, 066\,701,
  \doi{10.1088/0034-4885/74/6/066701}, 2011.

\bibitem[{Kosmopoulos et~al.(2021)Kosmopoulos, Kazadzis, Schmalwieser, Raptis,
  Papachristopoulou, Fountoulakis, Masoom, Bais, Bilbao, Blumthaler, Kreuter,
  Siani, Eleftheratos, Topaloglou, Gr\"obner, Johnsen, Svendby, Vilaplana,
  Doppler, Webb, Khazova, De~Backer, Heikkil\"a, Lakkala, Jaroslawski, Meleti,
  Di\'emoz, H\"ulsen, Klotz, Rimmer, and Kontoes}]{Kosmopoulos21}
Kosmopoulos, P.~G., Kazadzis, S., Schmalwieser, A.~W., Raptis, P.~I.,
  Papachristopoulou, K., Fountoulakis, I., Masoom, A., Bais, A.~F., Bilbao, J.,
  Blumthaler, M., Kreuter, A., Siani, A.~M., Eleftheratos, K., Topaloglou, C.,
  Gr\"obner, J., Johnsen, B., Svendby, T.~M., Vilaplana, J.~M., Doppler, L.,
  Webb, A.~R., Khazova, M., De~Backer, H., Heikkil\"a, A., Lakkala, K.,
  Jaroslawski, J., Meleti, C., Di\'emoz, H., H\"ulsen, G., Klotz, B., Rimmer,
  J., and Kontoes, C.: Real-time UV index retrieval in Europe using Earth
  observation-based techniques: system description and quality assessment,
  Atmospheric Measurement Techniques, 14, 5657--5699,
  \doi{10.5194/amt-14-5657-2021}, 2021.

\bibitem[{Lucas et~al.(2006)Lucas, McMichael, Smith, and Armstrong}]{Lucas2006}
Lucas, R., McMichael, T., Smith, W., and Armstrong, B.: Solar Ultraviolet
  Radiation: Global burden of disease from solar ultraviolet radiation, WHO
  Environmental Burden of Disease Series, 13, 2006.

\bibitem[{Mayer and Kylling(2005)}]{libradtran1}
Mayer, B. and Kylling, A.: Technical note: The libRadtran software package for
  radiative transfer calculations - description and examples of use,
  Atmospheric Chemistry and Physics, 5, 1855--1877,
  \doi{10.5194/acp-5-1855-2005}, 2005.

\bibitem[{Meng et~al.(2013)Meng, Zhijun, and Borders}]{Meng2013}
Meng, Q., Zhijun, L., and Borders, B.~E.: Assessment of regression kriging for
  spatial interpolation – comparisons of seven GIS interpolation methods,
  artography and Geographic Information Science, 40, 28--39,
  \doi{110.1080/15230406.2013.762138}, 2013.

\bibitem[{Oliver and Webster(1990)}]{Oliver1990}
Oliver, M.~A. and Webster, R.: Kriging: a method of interpolation for
  geographical information systems, International journal of geographical
  information systems, 4, 313--332, \doi{10.1080/02693799008941549}, 1990.

\bibitem[{Pfister et~al.(2003)Pfister, McKenzie, Liley, Thomas, Forgan, and
  Long}]{PfisterEA03}
Pfister, G., McKenzie, R.~L., Liley, J.~B., Thomas, A., Forgan, B.~W., and
  Long, C.~N.: Cloud Coverage Based on All-Sky Imaging and Its Impact on
  Surface Solar Irradiance, Journal of Applied Meteorology, 42, 1421 -- 1434,
  \doi{10.1175/1520-0450(2003)042<1421:CCBOAI>2.0.CO;2}, 2003.

\bibitem[{Schallhart et~al.(2008)Schallhart, Blumthaler, Schreder, and
  Verdebout}]{Schallhart08}
Schallhart, B., Blumthaler, M., Schreder, J., and Verdebout, J.: A method to
  generate near real time UV-Index maps of Austria, Atmospheric Chemistry and
  Physics, 8, 7483--7491, \doi{10.5194/acp-8-7483-2008}, 2008.

\bibitem[{Schenzinger et~al.(2023)Schenzinger, Kreuter, Klotz, Schwarzmann, and
  Gr\"obner}]{amt-2023-188}
Schenzinger, V., Kreuter, A., Klotz, B., Schwarzmann, M., and Gr\"obner, J.: On
  the production and validation of satellite based UV index maps, Atmospheric
  Measurement Techniques Discussions, 2023, 1--26, \doi{10.5194/amt-2023-188},
  2023.

\bibitem[{Schmalwieser and Schauberger(2001)}]{Schmalwieser2001}
Schmalwieser, A.~W. and Schauberger, G.: A monitoring network for
  erythemally-effective solar ultraviolet radiation in Austria: determination
  of the measuring sites and visualisation of the spatial distribution,
  Theoretical and Applied Climatology, 69, 221--229,
  \doi{10.1007/s007040170027}, 2001.

\bibitem[{Verdebout(2000)}]{Verdebout2000}
Verdebout, J.: A method to generate surface {UV} radiation maps over Europe
  using {GOME}, Meteosat, and ancillary geophysical data, Journal of
  Geophysical Research, 105, 5049--5058, 2000.

\bibitem[{Webb and Engelsen(2006)}]{webb2006}
Webb, A.~R. and Engelsen, O.: Calculated Ultraviolet Exposure Levels for a
  Healthy Vitamin D Status, Photochem. Photobiol., 82, 1697,
  \urlprefix\url{https://doi.org/10.1111/j.1751-1097.2006.tb09833.x}, 2006.

\bibitem[{Webb et~al.(2011)Webb, Kift, Berry, and Rhodes}]{Webb2011}
Webb, A.~R., Kift, R., Berry, J.~L., and Rhodes, L.~E.: The vitamin D debate:
  Translating controlled experiments into reality for human sun exposure times,
  Photobiol., 87, 741–745, \doi{10.1111/j.1751-1097.2011.00898.x}, 2011.

\bibitem[{WHO/WMO(2002)}]{who2002}
WHO/WMO: Global solar UV index : a practical guide, World Health Organization,
  a joint recommendation of the World Health Organization, World Meteorological
  Organization, United Nations Environment Programme, and the International
  Commission on Non-Ionizing Radiation Protection, 2002.

\end{thebibliography}

\end{document}